\documentstyle[preprint,revtex]{aps}
\tightenlines

\begin{document}
\draft

\begin{title}
Ordering and finite-size effects in the dynamics of one-dimensional
transient patterns
\end{title}

\author{A. Amengual, E. Hern\'andez-Garc\'\i a, and M. San Miguel}

\begin{instit}
Departament de F\'{\i}sica \\
Universitat de les Illes Balears\\
E-07071 Palma de Mallorca (Spain)
\end{instit}

\begin{abstract}
We introduce and analyze a general one-dimensional model for the description
of transient patterns which occur in the evolution between two
spatially homogeneous states. This phenomenon occurs, for example,
during the Fr\'eedericksz transition in nematic liquid crystals.
The dynamics leads to
the emergence of finite domains which are locally periodic and
independent of each other. This picture is substantiated by a finite-size
scaling law for
the structure factor. The mechanism of evolution
towards the final homogeneous state is by local roll destruction and
associated reduction of local wavenumber. The scaling law breaks down for
systems of size comparable to the size of the locally periodic domains. For
systems of this size or smaller, an apparent nonlinear selection of a global
wavelength holds, giving rise to long lived periodic configurations which do
not occur for large systems. We also make explicit the unsuitability of a
description of transient pattern dynamics in terms of a few Fourier mode
amplitudes, even for small systems with a few linearly unstable modes.
\end{abstract}


\narrowtext

\section{INTRODUCTION}

The most simple situation considered in the context of pattern formation
studies \cite{Newell} is the one in which an homogeneous stable steady state of
a system becomes unstable at a threshold value of a control parameter, so
that beyond threshold the new stable state is time independent and with a
well defined spatial periodicity. In such situation it is first generally aimed
to describe some static properties, such as the threshold value of the control
parameter, possible wavelengths of the pattern and possible higher-order
bifurcations. Other set of interesting questions is associated with the
transient  dynamics of the pattern formation process given an initial unstable
homogeneous state \cite{SHours}. A different physical situation is the one of
transient pattern formation during the temporal evolution between two
homogeneous
steady states: In this case an homogeneous stable steady state becomes unstable
beyond a given  threshold and the transient evolution to the final
homogeneous  stable steady state involves a process of pattern
growth and decay. Most of the usual mathematical techniques used
to describe the first situation seem to fail for problems involving
transient pattern dynamics.

Transient pattern formation is well documented from the experimental
point of view for different instabilities in nematic liquid crystals
\cite{Cargese,Winkler}. A typical situation is the magnetic Fr\'eedericksz
transition in which for a large enough applied magnetic field the nematic
director does not reorientate homogeneously but a striped pattern with a
characteristic wavelength emerges. This pattern can last from minutes to hours
depending on the specific geometry and the material, but it finally
disappears leading to the homogeneously-reoriented final equilibrium state.
For most geometries of the system, the pattern admits a good description by
one-dimensional models. A good  summary of
different situations considered in this context is given in Ref.
\cite{Winkler}.
Generally speaking the transient pattern is associated with the coupling of the
director field with hydrodynamic variables so that the fastest response
to the applied field is not homogeneous in space \cite{Guyon}.
A mechanism of wavelength selection based on this idea of fastest response has
been proposed \cite{Cargese,Winkler,Guyon,SMS86,Kramer}: the well-defined
observed periodicity has been associated with the mode of fastest growth and
its
dependence with the applied magnetic field has been considered theoretically
and
experimentally. However, a nonlinear mechanism of wavelength selection has also
been invoked and substantiated by some experimental and restricted numerical
studies \cite{Srajer}. The process of pattern formation in nematics can
be described in detail through the full set of nematodynamic equations
\cite{Gennes} which have been consistently formulated also in the presence of
thermal fluctuations \cite{SMS86}. Such equations have been discussed in a
variety of situations \cite{Cargese,Winkler,Kramer,Srajer}. These detailed
discussions of a rather complicated set of equations might hide some general
features of the problem of transient pattern formation and development. Our aim
in this paper is to propose and analyze  a generic model for one-dimensional
transient pattern formation which describes general features of these problems.
It is obvious that a precise comparison with experiment would require the
consideration of some specific details. Nevertheless, we hope that general
aspects as wavelength selection (if such selection does occur), mechanisms of
pattern evolution and the important issue of finite size effects can be
understood from this general model.

The model we analyze is defined by the following equation for
a scalar variable $\theta (x,t)$:
\begin{equation}
\dot \theta (x,t) = (a - {\partial_x}^2)
({\partial_x}^2 \theta + c \theta - b \theta^3) \ \ .
\label{unaecu}
\end{equation}

\noindent The dot denotes temporal derivative. The linearized version of
this equation in Fourier space  \begin{equation}
\dot \theta_q = \omega(q) \theta_q
\end{equation}

\noindent involves an amplifying factor

\begin{equation}
\omega(q) = (a + q^2) (c - q^2) \ \ .
\label{factorw}
\end{equation}

\noindent As seen in Fig. 1, this factor is such that for $c>0$
there is a range of unstable modes for which $\omega (q)  > 0$.
If $a < 0$ the range of unstable modes does not include the
mode $q = 0$ and the linear regime is qualitatively the same as
for the well known Swift-Hohenberg equation \cite{sh} used to describe the
formation of stationary patterns. We are here interested in the situation in
which $a > 0$ for which the range of unstable modes is $(-q_c,q_c)$, with $q_c
\equiv c^{1/2}$. If in addition, $a<c$, the mode of fastest growth  becomes
different from zero: $q_m = ( (c-a) / 2 )^{1/2}$. This implies the existence of
an instability that for $0<a<c$ involves the  linear growth of a pattern with
a characteristic periodicity given by $q_m$. Throughout the paper we fix
$c \equiv 1$ and study (\ref{unaecu}) with periodic boundary conditions.

Equation (\ref{unaecu}) can be motivated as an approximation to the
complete nematodynamic equations describing the magnetic Fr\'eedericksz
transition in a nematic. Under ordinary assumptions for the twist
geometry \cite{Gennes} one obtains coupled equations for the angle orientation
$\theta(x)$ of the director and a component of the velocity field, both
in a plane in the middle of the sample. The approximation of negligible
inertia permits the elimination of the velocity variable. This gives
rise to an effective wavenumber-dependent viscosity which appears as
a $q$-dependent kinetic coefficient \cite{SMS86}. For small deformations
$\theta$ and in the small wavenumber limit \cite{maxilanger,Caceres} one
recovers
equation (\ref{unaecu}), where the factor $(a - {\partial_x}^2)$ is the
remaining part of the effective viscosity \cite{foot1}.

Within the context of model (\ref{unaecu}), we address in this paper three
general questions associated with transient patterns dynamics in the
intermediate nonlinear regime after the initial pattern emergence and before
the late stages in which it disappears.
The first question is the domain of validity
of the linear theory and the possible existence of a wavelength-selection
principle in the nonlinear regime. The second question is the characterization
of the mechanism governing pattern evolution. Thirdly we ~examine finite-size
effects which could be preponderant in the question of a nonlinear
selection principle. Our results indicate that for large systems a
wavelength is initially selected in the linear regime but, on the
average, it then changes monotonically in time approaching a final
homogeneous state. During this evolution the system can be described as
composed of several regions which evolve independently of each
other. A local mechanism of roll destruction operates in such regime.
This description is substantiated by a scaling law for the structure
factor of systems of different size which reveals the presence of
uncorrelated regions of a characteristic length. This length is essentially
time-independent in the regime of dynamical evolution examined. For
systems of size comparable or smaller than this length an apparent
nonlinear selection of a wavelength occurs: configurations of well
defined periodicity, which is not the one linearly selected, last for very
long times, before evolving to the final homogeneous state.
The reason why such periodic configurations are long lived is that they
correspond to unstable stationary solutions which, for small system sizes,
are approached in the initial transient regime. For large system sizes
such configurations are not approached in the dynamical evolution from a
typical initial configuration.
A characteristic of small systems is that a very
small number of modes are linearly unstable at $t=0$ so that a description in
terms of coupled ordinary differential equations for the amplitude of a few
modes seems natural. We show that such description can give qualitatively wrong
results. The basic simple reason
for this fact is that such modes are not stabilized by nonlinear terms in
the full equation, because they are not associated with stable solutions.

We finally wish to mention the relation of our analysis with another
well-known physical situation in which transient patterns occur, namely
spinodal decomposition \cite{Sahni}.
The process of phase separation
of a binary mixture quenched to a temperature below the critical
temperature also displays a transient pattern with a time dependent
characteristic length. This analogy was pointed out many times \cite{Cargese}.
However, in addition of differences in time scales \cite{SMS86} there is
another
important difference between this problem and the one of transient patterns in
nematics and it is that the dynamics of spinodal decomposition is constrained
by
a conservation law. The fact that $a \not = 0$ in (\ref{unaecu}) indicates that
there is no conservation law for the spatial integral of $\theta(x,t)$.
 One-dimensional ($d=1$) spinodal decomposition is
physically artificial and consequently has not been studied in great detail
from the physical point of
view. The initial linear regime (Cahn-Hilliard-Cook theory \cite{Sahni}) and
the
very late stages of phase separation \cite{Langer} are well understood,
but no detailed study seems available for the $d=1$ intermediate
nonlinear regime in which we
are here interested. The initial \cite{SMS86,Arias} and very late stages
\cite{maxilanger} of pattern dynamics in the magnetic Fr\'eedericksz transition
have been already studied in some detail by analogy with the problem of
spinodal
decomposition. Standard nonlinear theories of spinodal decomposition such as
that
of Langer, Bar-on, and  Miller \cite{LBM} do not seem applicable in $d=1$:
since
they are based on the competition between locally ordered equivalent stable
states  they do not include the competition between different wavenumbers
\cite{Grant} which seems essential in $d=1$.
The other classical problem of the dynamics of phase transitions,
namely the dynamics of an order-disorder transition \cite{Sahni} does not
involve a conservation law but it neither has a  periodicity  selected by the
dynamics of the process.

The outline of this paper is as follows. In section 2, we discuss the main
characteristics of the model given by (\ref{unaecu}) and its stationary
configurations, then we compare it with other related models. Section 3
describes
our numerical results for large systems. Finally the description of a small
system in terms of the amplitudes of a few modes is discussed in section 4.
Throughout the paper we restrict ourselves to $d=1$ and we neglect thermal
fluctuations. The role of fluctuations and two dimensional effects will be
analyzed elsewhere.

\section{A MODEL FOR GENERIC ASPECTS OF TRANSIENT PATTERN FORMATION}

A first important property of the model given by equation (\ref{unaecu}) is
that
it can be  written in a potential form
\begin{equation} \label{potential}
\dot\theta(x) = \Gamma {\delta F[\theta] \over
                                           \delta \theta(x) }
\end{equation}
where the kinetic coefficient $\Gamma$ is the operator
\begin{equation} \label{kinetic}
\Gamma = a - {\partial_x}^2 \ ,
\end{equation}
and $F[\theta]$ is
\begin{equation} \label{freeenergy}
F[\theta] = \int dx\ \left[ -{1 \over 2}\theta(x)^2 + {b \over 4}\theta(x)^4  +
{1 \over 2}|\partial_x \theta(x)|^2 \right]  \ \ .
\end{equation}
{}From (\ref{unaecu}) we can show that $F$ is a good Lyapunov functional, in
the  sense that $dF/dt \le 0$:
\begin{equation} \label{fdrops}
{ dF[\theta] \over dt } = \int dx {\delta F[\theta] \over
                                 \delta \theta (x)} \dot\theta(x) =
- \int dx {\delta F[\theta] \over \delta \theta (x)} \Gamma
              {\delta F[\theta] \over \delta \theta (x)} \le 0    \ \ .
\end{equation}
The last inequality holds because $\Gamma$ is a positive definite
operator (if $a>0$), as can be seen from its expression in Fourier space.

The picture of the evolution is then that the system evolves, continuously
decreasing $F[\theta]$, until a minimum of $F[\theta]$ is found, thus stopping
the evolution. Note that such minima, stationary solutions of (\ref{unaecu}),
are independent of $\Gamma$ and of the parameter $a$ and they are solutions
of  the simpler equation
\begin{equation}  \label{modela}
{\partial_x}^2\theta(x) + \theta(x) - b \theta(x)^3 = 0 \ \ .
\end{equation}
An independent demonstration of the fact that with periodic boundary conditions
the only stationary solutions of (\ref{unaecu}) are those of (\ref{modela}) can
be set up by writing (\ref{unaecu}) with $\dot\theta(x)=0$ as the set of two
equations:

\begin{equation} \label{modelay}
{\partial_x}^2\theta(x) + \theta(x) - b \theta(x)^3 = y(x)
\end{equation}
\begin{equation}\label{gammay}
\Gamma y(x) = 0 \ \ .
\end{equation}
Eq. (10) is a linear second order ordinary differential equation whose
general solution is a linear combination of two exponentials. Only the
combination leading to $y (x ) = 0$ satisfies periodic boundary
conditions,  so that (\ref{modelay}) reduces to  (\ref{modela}).
The qualitative features of all the stationary solutions of (\ref{unaecu}) can
be discussed by writing (\ref{modela}) as \cite{pot}
\begin{equation}  \label{newton}
{d^2\theta(x) \over dx^2} = -{dV(\theta) \over d\theta}
\end{equation}
which resembles a Newton equation for the motion of a particle of unit mass
in a potential
\begin{equation} \label{V}
V(\theta) = {1 \over 2}\theta^2  - {b \over 4}\theta^4   \ \ ,
\end{equation}
the role of `time' being played by the coordinate $x$. From this analogy, the
bounded solutions of (\ref{modela}) can be classified in three types:
\begin{itemize}
\item[a)]
 The uniform solutions $\theta(x)=1/\sqrt b$ and
$\theta(x)=-1/\sqrt b$.
\item[b)]
 The uniform solution $\theta(x)=0$.
\item[c)]
 A family of solutions represented by nonlinear oscillations in the
potential $V(\theta)$. The maximum possible frequency corresponds to
oscillations of small amplitude around $\theta=0$, being its period $2\pi$.
The minimum frequency corresponds to trajectories in which $\theta(x)$ remains
mostly near $\pm 1/\sqrt b$, with short excursions (domain walls) linking
both states.
In summary: there are periodic solutions to
(\ref{modela}), which we denote by $\psi_q(x)$, each one containing a
different fundamental wavenumber $q$ and its harmonics, with $0<q<1$.
\end{itemize}

An important question is the linear stability of such stationary solutions.
In order to consider this question, Eq.(\ref{modela}) does not contain enough
information and the full dynamical equation (\ref{unaecu}),
linearized around the stationary solution being checked, is needed.
Linearization around the uniform solutions is immediate and it is found that
the solution $\theta=0$ is linearly unstable, and both $\theta = 1/\sqrt b$
and $\theta = - 1/\sqrt b$ are linearly stable. These are also the absolute
minima of the functional $F[\theta]$, so that they represent the stable
equilibrium phases. The analysis of the stationary periodic
solution $\psi_q(x)$, of fundamental wavenumber $q$, is performed with the
introduction of $\theta(x,t) \equiv \psi_q(x) + \Delta(x,t)$ and
linearization in $\Delta$. The resulting equation is
\begin{equation} \label{floquet}
\dot \Delta(x,t) = \Gamma \left[1-3b\psi_q(x)^2+{\partial_x}^2 \right]
\Delta(x,t)     \ \ .
\end{equation}
The general analysis of this linear equation with periodic coefficients
requires of Bloch or Floquet
theory \cite{bloch}. A simplified situation was considered in \cite{maxilanger}
for the case in which $q$ is small, so that the solution consisted basically
of domains
of the stable phases separated by thin domain walls. In that case it was found
that the periodic solutions were linearly unstable.
It can be generally shown that all
the periodic solutions are unstable by studying its stability with respect to
a uniform perturbation. The argument is as follows: Let us introduce the
initial
perturbation $\Delta(x,t=0)=\Delta_0, \ \forall x$, and consider
the initial time $t=0^+$, when $\Delta(x,t) \approx \Delta_0$. Let be
$x \equiv 0$ one of the places in which $\psi_q=0$. Near $x=0$,
$\psi_q(x)^2$ will have a positive parabolic shape, so that
$\partial_x^2[\psi_q(x \approx 0)^2] > 0$. Introducing this in
Eq.(\ref{floquet})  we find that
${\rm sign}[\dot\Delta(x \approx 0,t=0^+)] = {\rm sign}[\Delta_0]$,
showing the instability of $\psi_q$ because a uniform initial perturbation
grows. The consequence of the instability of {\sl all}
the periodic solutions is that  none of them can represent the final state of
the
evolution, as long as a  non-zero amplitude is given in the initial condition
to
the mode with wavenumber $q=0$, or if  noise is present in the system.

After this summary of the general properties of Eq.(\ref{unaecu}) it is
interesting to compare them with the properties of other related models
studied in the literature. To this end we write (\ref{unaecu}) (with
$c=1$) as \begin{equation} \label{desarrollada} \dot \theta =
-{\partial_x}^4 \theta + (a - 1) {\partial_x}^2 \theta +
 a \theta - a b \theta^3 + b {\partial_x}^2 \theta^3      \ \ .
\end{equation}

For $0<a<1$, the uniform solution $\theta = 0$ is unstable and the linear
analysis predicts the growth of modes with wavenumber $q \not = 0$. It is then
natural to relate this equation with others for which a periodic pattern grows
from the unstable uniform solution. An archetypal example
of such equations is the Swift-Hohenberg equation \cite{sh}
\begin{equation} \label{sh}
\dot \theta =
\left[\gamma^2 - (1 + {\partial_x}^2) ^2 \right] \theta - b\theta^3    \ \ .
\end{equation}
The linear stability analysis of this equation leads in fact to the same linear
growth spectrum than Eq.(\ref{unaecu}) except  for an important difference in
sign in the regions around $q \sim 0$: the modes in
this region slowly  grow in our model $(a > 0)$ whereas they are
damped in the Swift-Hohenberg case (due to  the fact that $\gamma^2$ is
positive). Other aspects of the initial stages in pattern development are
qualitatively similar for both models. Another more fundamental difference is
that the Swift-Hohenberg equation admits a family of  {\sl stable} periodic
solutions, one of which will give the final state of the  evolution, whereas
(\ref{unaecu}) admits no other stable solutions than the  uniform
$\theta=\pm 1/\sqrt b$. This difference comes from a combination of  the
different sign of $\gamma^2$ {\sl versus} $-a$ and of the additional term
${\partial_x}^2 \theta(x)^3$ in (\ref{desarrollada}). Thus, the evolution at
late  times will be completely different in both cases.

These differences have important methodological consequences: in order
to  study pattern formation in cases exemplified by the Swift-Hohenberg
equation, a common  strategy is to take as a small parameter the range of
unstable modes around the  most unstable one, which is small near a
bifurcation point, and then  obtain a nonlinear equation for the amplitude of
the most unstable mode. The  form of this equation is greatly determined by
the symmetries of the problem,  and by the assumption of being the first step
in a uniform expansion. This  strategy can not be applied to our problem,
because the characteristic shape  of the linear instability spectrum in
(\ref{factorw}) with a fastest growing mode $q_m \not = 0$ is only obtained
for $c > a$ which is far enough from the  bifurcation point $(c = 0)$, and the
band of unstable wavenumbers is as large as the wavenumber of the most unstable
mode (because the mode with $q=0$ has to be included in the description). Then
bifurcation theory and normal forms are of no much  help in our problem. In
addition, the fact that the mode of fastest growth is not associated with a
stable solution precludes the use of approximations based in the saturation of
the linearly fastest growing mode, such as those in \cite{kyg&co}.

Another class of models with which it is natural to compare our model is
the one represented by the Fisher-Kolmogorov equation, also known as
Ginzburg-Landau equation for a real variable, or, with an added noise term,
model A of critical dynamics \cite{Sahni,hh}:
\begin{equation} \label{timemodela}
\dot \theta(x,t) = {\partial_x}^2 \theta(x) +
\theta(x) - b \theta(x)^3   \ \ .
\end{equation}
The stationary solutions of this equation are exactly the same as in our model,
and the only stable solutions are, as in our case, the uniform
$\theta=\pm 1/\sqrt b$. In fact, the analysis in \cite{maxilanger} shows that
at very long times, the dynamics of the domain walls in (\ref{unaecu}) is the
same as in (\ref{timemodela}). The main difference is, however in the
conditions created by the initial linear instability if $a<1$. The fastest
growing mode
in (\ref{timemodela}) is the one with wavenumber $q=0$, which corresponds
also  to the final state. Then theories such as those in \cite{kyg&co} are good
descriptions  of the evolution for all times.

In some sense, the time evolution of our model is a crossover between a linear
behavior close to that in the Swift-Hohenberg model, and final stages similar
to those in
(\ref{timemodela}). Perhaps, this is why the closestly related model is the
one represented by the Cahn-Hilliard equation, describing spinodal
decomposition in binary mixtures and alloys. It is also known, when a noise
term is added, as model B of critical dynamics \cite{Sahni,hh}. Formally, this
model is obtained by putting $a=0$ in Eq.(\ref{unaecu}). In this case we
have also an initially periodic structure which coarsens in time to approach
$q=0$. The main difference with our model is that the spatial
integral of $\theta(x)$ is conserved by the Cahn-Hilliard equation, so that a
completely uniform solution  can not be approached unless
$\int dx\ \theta(x,t=0) = 0$. In the generic case, the final state is the
coexistence of two domains of the
stable phases and not only one as in our case. Thus,  the final stages of
evolution
should be very different in  both models  \cite{maxilanger}. The fact that, in
addition to the fastest  growing mode, the mode with $q=0$ is also linearly
unstable in (\ref{modela}) implies a wide range of unstable  wavenumbers in the
initial  regime, leading to a wide
spectrum during the nonlinear stages. Time-dependent configurations do not
approach closely to any of the unstable periodic solutions. The consequence
is that theories such as that  of Langer \cite{Langer} assuming that the
system is
close to one of the  stationary unstable solutions at each time, will be only
of
certain usefulness at extremely long times \cite{maxilanger}, where the mode
with $q=0$ will be the dominant one.

\eject
\section{NUMERICAL STUDY OF TRANSIENT DYNAMICS FOR LARGE SYSTEMS}

The time evolution of $\theta(x,t) \ $  from an initial condition close to the
unstable
steady state $\theta(x)=0$ has been calculated by solving numerically equation
(\ref{unaecu}) on a grid of $N$ points. In the remaining part of the paper,
we fix the value of the parameters in (\ref{unaecu}) to $a=0.002$, $b=3$ and
$c=1$ as appropriate for typical values of the parameters in the
nematodynamic equations \cite{Srajer,foot1}. In this case $q_m = 0.7$.
We have used a centered finite-difference scheme up to
order $(dx)^4$ to approximate the spatial derivatives. A predictor-corrector
method with one step has been used to determine $\theta$ at $t + dt$. A
suitable value for $dx$ has been determined integrating the equation in the
linear regime and comparing the growth rate of the unstable modes obtained
numerically with the one calculated analytically. A value of $dx= 0.25$ has
been
thereby chosen. The most unstable mode has a wavelength of $\lambda_m = 2 \pi /
q_m \approx 8.98$ which corresponds to approximately $36$ grid points. The
length of the system is $L = N dx$. We have considered a range of system sizes
from $ L = 64$ to $L = 256$  and we have always taken periodic boundary
conditions. The time step used is $10^{-4}$. For $dt$ larger than $0.004$
numerical instabilities were observed. For values of $dt$ ranging from $2 \cdot
10^{-4}$ to  $5 \cdot 10^{-5}$ the discrepancies in $\theta(x) \ $ after $5$
units of time of integration  were smaller than $10^{-7}$.

The initial condition is written as
\begin{equation}
\theta(x,0) = {\epsilon \over 2} + \epsilon \sum_{q_n}
sin  \left( q_n x + \phi_{q_n} \right)
\label{inicial}
\end{equation}

\noindent where the sum is over modes $q_n = 2 \pi n / L$ and it has been
usually
taken to  run only over the unstable modes $q_n < q_c$; $\phi_{q_n}$ is a
random
phase  and $\epsilon$ is a small amplitude assumed
equal for all of the modes and whose value was arbitrarily set equal to
$ 2 \cdot 10^{-4}$. To obtain different initial configurations of
$\theta(x) \ $ the set of
random phase shifts $\phi_{q_n}$ was changed but not the amplitude $\epsilon$.
Hence, when talking about a different initial condition we mean a different
set of random phase shifts.

In addition to our transient dynamics study, we have also examined the
existence of
periodic stationary solutions (type `c' in the previous section). Starting
with a configuration of the form $\theta(x) = \epsilon \> sin(q_n x)$
involving a
single mode $q_n < q_c$  with an amplitude $\epsilon = 2 \cdot 10^{-4}$, the
pattern develops  with the growth of its harmonics until a stationary pattern
is
obtained. We know that this pattern is unstable and the velocity of its decay
has been tested numerically by adding a small amplitude
to all the modes with $q$ smaller than $q_c$. The decay was found to be always
extraordinarily slow.  This means that these unstable stationary solutions can
be long lived. When the initial condition (\ref{inicial}) is used, the pattern
develops in a way that none of these periodic stationary patterns is approached
during the transient dynamics provided the system size is large enough (see
next
section).

Our results for the transient dynamics study are summarized in figs. 2, 3 and
4 for the evolution of the configuration $ \theta (x, t)$, the associated
structure factor and the number of rolls of the pattern respectively. General
 features of the time evolution which manifest themselves in specific ways
in these figures are the following: A linear and a nonlinear regime of
evolution can be clearly identified. In the linear regime the pattern is
formed. In the nonlinear regime no mechanism of wavelength selection occurs,
but domains of a characteristic size exist. These domains include several
rolls and evolve in time in a way essentially independent of each other.

In Fig. 2a, the time
evolution of the pattern $\theta(x) \ $ for a particular initial condition is
shown. The
periodicity related to the linearly most unstable mode becomes apparent on
$\theta(x) \ $ in the initial stages.  Afterwards, we observe that rolls
disappear continuously.  The presence of regions of different periodicities
can be
observed. For example, at $t=40$ around 7 of these regions are distinguished. A
systematic method of identifying such regions is to find the  power spectra of
small regions in the pattern, and identify the maxima in  these local spectra
with
a dominant local wavenumber \cite{SHours}.  Explicitly, we have multiplied the
configurations in Fig.2a times a Gaussian of  width $\sigma=10$, unit height,
and centered at $x$. Then we have calculated  the maxima in the power spectra
of
such localized patches as a function  of $x$.  Figure 2b shows the result of
one
of such analysis, at time $ t = 40$, identifying  an $x$-dependent dominant
wavenumber. At this time the average wavenumber has already deviated from the
linearly most unstable one ($q_m= 0.70$) so that one has entered the nonlinear
regime. In Fig.2a we can also see that, at each time, a roll is disappearing in
that region whose local dominant  wavenumber is largest. Note that the
disappearance of a roll occurs locally since this does not affect
other regions  (compare for example $\theta(x) \ $ at $t=100$, $200$ and
$300$).
At the longest times arrived, there are regions were $\theta$ is already  close
to the value of the uniform solution $\theta = \pm b^{-{1 \over 2}} \approx
0.58$. After this time, it is expected that the pattern evolves towards
configurations where regions of $\theta = b^{- 1 /2}$ and $\theta = - b^{-
1/2}$  coexist separated by walls. The separation between these walls will be
rather large and the pattern evolution should be described by the  theory in
\cite{maxilanger}.

The structure factor $S(q,t)$ associated with $\theta (x, t)$ is defined from
the discretized configuration $\{ \theta(x_n = n dx,t), n= 1,...,N \}$ as

\begin{equation}
\label{sq}
S(q,t) = {1 \over N} \left| \sum_n e^{i x_n q} \theta(x_n,t) \right|
\end{equation}

\noindent  The vertical bars denote the modulus of a complex number. Note that
a
normalization factor $1/N$ has been included. With this choice of normalization
$S(q)$ is independent of system size for a uniform configuration
$\theta(x_n,t)$. It is also independent of N during the linear regime. The
allowed values of $q$ are of the form $n \> dq$ with $n$ an  integer between
$-N/2$ and $N/2$ and $dq= 2 \pi / L$. Figure 3a shows the time  evolution of
$S(q)$ averaged over $50$ independent initial conditions of the  form
(\ref{inicial}). Figure 3b shows the time evolution of the area $A(t) \equiv
\int  dq S(q,t)$ for  $4$ of these  independent initial conditions. Since
linear
theory predicts that the structure factor is  independent of the initial phases
$\phi_{q_n}$, the time at which the different curves in Fig. 3b begin to
separate signals the end of the linear regime. This happens  around $t \sim 20
$. During the linear regime the structure factor in Fig. 3a is shown to grow
with a maximum around the linearly most unstable mode $q_m = 0.7$.

After the linear regime, and reflecting a continuous elimination of rolls,
the maximum of $S(q,t)$
shifts towards small $q$'s as times goes on (see Fig. 3a).
This continuous drift characterizes the elimination of rolls and it
eliminates the idea of a nonlinearly selected wavelength. During the
nonlinear regime the structure factor develops important contributions for
short and long wavelengths which indicate the existence of strong competition
between many modes. The existence of different domains in the configuration
$\theta (x, t)$ can be characterized from the structure factor by defining a
correlation length $l_c = 2 \pi / w$, where $w$ is one half of the width of the
peak of $S(q,t)$ at its mean height. This  length has been plotted in
Fig. 3c for a system of size $L=128$. It is around  $1/3$ of the system
length for this system size. Hence, we talk about approximately $3$
uncorrelated
zones in the system which evolve independently. The validity of this view is
enhanced by the evolution observed in systems with $L=256$, as discussed
above, and also with $L=64$: $w$ turns out to be roughly independent of system
size and time (during the nonlinear regime we consider here).
The domain size is probably
determined by the interplay between the intensity $\epsilon$ of the initial
condition and the shape of $\omega(q)$.

Figure 4 shows the average number of zeros per unit length of $\theta(x) \ $
as a function of time. This quantity identifies the
number of rolls per unit length of the pattern and it measures the `average'
wavenumber in the system. The inset in this figure represents the average
over initial conditions of  the number of zeros per unit length {\sl vs} a mean
wavenumber defined as $\langle q \rangle \ $ $\equiv \int q S(q,t) dq / A(t)$.
The upper part of the curve in the inset corresponds to the very early
initial regime. Beyond this regime the number of zeros is linearly
related to $\langle q \rangle \ $ so that a description in terms of any of
these
quantities is equivalent. The number of zeros per unit length is seen to
grow during the linear regime of pattern emergence reaching a value
around 0.22 which corresponds to the global wavenumber selected by
fastest linear growth, $q_m = 0.7$. This number of zeros remains constant
for a little while beyond the end of the linear regime at $ t \sim 20$.
Later the average number of zeros decreases
monotonically in time making again clear that we can not identify any
nonlinearly selected wavenumber. This result is different to the one
found in Ref. 8 for a similar system. We will show in the next section
that an apparent wavelength selection might occur due to finite size
effects. We have attempted to obtain a growth law for the decay process
in Fig. 4. The decay is certainly slower than the one
given by a power law. A relation $\langle q \rangle \ $ $= 21.3 - 1.27 \ln t$
seems to fit
the curve for $t \geq 90$. Nevertheless, a decay $(\ln t)^\beta$ could also
be fitted. Determining the exact law and an accurate  exponent will need of
a considerable increase in the statistics and it is  not our goal here. We
note that a logarithmic decay is the one expected  from one-dimensional
domain-wall dynamics \cite{maxilanger,Langer}, but the  largest times in our
simulations are still far from the regime in which the  pattern is composed
of domains of the equilibrium phases separated by  thin walls.

An important question in relation with the existence of domains of a
characteristic size is the dependence of the transient dynamics on the
system size $L$. It is already seen in Fig. 4 that the evolution of the
number of zeros per unit length is essentially the same for two system
sizes. We have checked that the time scales of evolution are independent
of system size for $L \,\vcenter{\hbox{$\buildrel\textstyle>\over\sim$}}\,$
50.  Taking advantage of this fact, a systematic study of the dependence on
$L$ of the temporal evolution of the structure factor can be carried out by the
analysis of the evolution of the area $A$ of $S(q,t)$. This area is shown in
Fig. 5a for systems of different size. From $t=75$ to the longest time
considered ($t=200$), a relationship of the  form $A(L,t)=L^\alpha f(t)$ is
satisfied, with $\alpha= -0.49 \pm 0.01 \approx  1/2$, for system sizes large
enough. Figure 5b shows the validity of this  scaling behavior.  The
combination
of this result with the existence of size-independent time-scales of evolution
implies that the dependence of the  structure factor on system size factorizes
out: \begin{equation} \label{scaling} S(q,t,L) = L^{-{1 \over 2}} F(q,t)
\end{equation} It should be stressed that this factorization is not a trivial
consequence of the normalization factor in (\ref{sq}). This
finite-size scaling form contrasts with the one found in the dynamics of
order-disorder transitions and of spinodal decomposition \cite{FSS}  in $d
\ge 2$, where the scaling function $F$ depends on system size as
$F(qL,t/L^z)$, and the exponent  $\alpha$ is $0$ instead of $-1/2$
\cite{footsq}. The
reason for these differences can be  elucidated by noting that the exponent
$\alpha=-1/2$ is a manifestation of the fact that the system  is
composed of uncorrelated regions of a size independent of the system
size. This can be seen from the following argument: If the system is made of
many uncorrelated regions, the
law of large numbers implies that the sum in (\ref{sq}) approaches a
Gaussian variable of standard deviation
proportional to the square root of the number of independent zones. Since
the size of the zones is independent of $L$, this standard deviation is
proportional to  $L^{1/2}$. The average of
the modulus of a complex Gaussian variable is proportional to its standard
deviation, and the normalization factor
$1/N  = dx/L$ present in definition (\ref{sq}) completes the factor
$L^{-1/2}$ in (\ref{scaling}). This argument
confirms  again the important dynamical role of the `domains of different
wavenumber' identified before. To further stablish this picture, we note
that the law $A \sim L^{-1/2}$ should fail when there is only one or less
than one domain in our system. According to the results of Fig. 3c, this
will happen for $L \le 50$ ($N \le 200$). This has been  checked with a
system of size $L=32$ ($N=128$) so that it is well described by  a single
local wavenumber during the time interval included in Fig. 5b. As seen in
that figure, this system does not fit into the finite-size-scaling
description. A discussion of the dynamics of these small-size systems is
given in the next section.

In the dynamics of conventional phase transitions such as order-disorder
transitions or spinodal
decomposition in $d \ge 2$ one can also identify independent domains
containing some
`ordered phase'. The difference with our model is that in the usual case such
`ordered phase' is close to an equilibrium stable phase. Then domains growth
and coalesce until one (or two in the case of spinodal decomposition) of them
reaches the size of the whole system. This domain growth and saturation
produces the $t/L^z$ dependence of the scaling function in time. In our case,
the domains contain an `unstable periodic phase', so that there is no driving
force for growth, and the domains keep its size roughly constant
(see Fig. 3c), but continuously reduce their wavenumber by local roll
destruction. This  process will presumably continue until each domain
contains only one of the  uniform stable phases $\pm b^{-1/2}$ and then a
growth
of the domains  similar to that in the model (\ref{timemodela}), as
described in \cite{maxilanger}, is expected.

\section{SMALL SYSTEMS}

Following our discussion above, we will call a system `small'
when its size is smaller than the correlation length discussed in the
previous  section ($l_c \sim 50$), so that it contains effectively only one
domain  of quite uniform wavenumber. In this context it is worth reminding here
how different characteristics of a system reflect  in their Fourier
description. First, for a small system, modes are more separated than in a
larger system ($dq = 2 \pi / L$).  Second, the difference
between a discrete or a continuous (in $x$) system is that in the first case
there is a minimum length, $dx$, so that there appears a maximum wavenumber
$2 \pi/dx$ and only a finite number of modes is at play if $L$ is finite. In
the continuum case ($dx \rightarrow 0$), the wavenumber
cut-off goes to infinity and we have an infinity of modes, but their
separation continues to be determined by $L$. As usually recognized, a
physically continuous system admits a discrete  description of minimum
length $dx$ when $2 \pi /dx$ is larger than any wavenumber relevant to the
evolution of the system under study.

To study the behavior of `small systems', we have performed numerical
integrations of equation (\ref{unaecu}) for a range of values of $L$ for
which  only three modes are linearly unstable at t=0: $q_n=n dq, \>
n=0,1,2$. The  mode $q_2$ is the most unstable one. The explicit results for
the amplitude of
the three unstable modes and the first linearly stable mode for the case $L=18$
$(N=72, dq \approx 0.349)$ are shown in Fig. 6a as obtained from the direct
numerical integration of (\ref{unaecu}) with the initial condition
(\ref{inicial}). The results indicate
that the mode  $q_1$ will dominate the pattern for a
long time after an initial short
regime in which the fastest growing mode $q_2$ dominates.
In such small systems, a well defined periodicity is observed during a long
time interval, so that an apparent nonlinear selection of the mode $q_1$
occurs.
This happens because the stationary  unstable solution $\psi_{q=q_1}$,
dominated by a mode and its  harmonics, is closely approached during the
evolution, in contrast with what we obtain for larger systems. Such
configuration
is, nevertheless, unstable. The pattern finally decays and the configuration
becomes space homogeneous, as expected. This happens at a time
$t \sim 1.6 \cdot 10^5 \> $. This time is not shown in
figure 6 and should be compared with the time scale in that figure. For such
small systems with only three linearly unstable modes, and given that the
linearly stable modes have a very small amplitude during the whole time
evolution, it is natural trying to describe the dynamics in terms of three
coupled ordinary equations for the complex amplitudes $\theta_0$,
$\theta_{q_1}$
and $\theta_{q_2}$, all the other amplitudes assumed to be zero. Truncation to
a
small number of modes is a  largely used technique in the literature
\cite{modos3}. Equation (\ref{unaecu}) is written in Fourier space including
only the modes $q_n = n dq$, with $n = 0, 1$ and $2$ as

\begin{equation}
\dot \theta_{q_n} =
\omega(q_n) \theta_{q_n} - b (a + {q_n}^2 )
\sum_{i,j=0,1,2} \theta_{q_i}  \theta_{q_j}  \theta_{q_n - q_i - q_j}
\label{enFourier}
\end{equation}

\noindent The dependence on system size appears through the values
of $q_n$, which depend on $dq$.

Equation (\ref{enFourier}) presents stationary solutions whose stability
properties are different from the corresponding solutions of
(\ref{unaecu}). For example,  $\theta_{q_2}= ((1- q_2^2)/(3b))^{1/2}$ and all
the other amplitudes equal to  zero is a stationary solution of
(\ref{enFourier}).
When the linear stability analysis around  this solution is carried out, the
result depends on the value of $dq$: The solution is stable against homogeneous
perturbations if  $dq < 8^{-1/2}=0.354$, for any value of $a$.
It is stable against perturbations of
wavenumber  $q_1$ only if $dq < 7^{-1/2}=0.378$
(value again independent of $a$). This behavior contrasts with
the exact properties of Eq.(\ref{unaecu}), well reproduced by its numerical
integration discussed above, for which there are no other
stable  solutions than the one associated to $q_0$, {\sl independently} of
the value of  $dq = 2\pi/L$. The differences in stability properties come from
the truncation to a small number of modes, equivalent to replacing the
spatially
continuous system by a discretized version. Such differences anticipate that a
description in terms of a few modes might be qualitatively incorrect. To
substantiate this point, we have numerically solved  Eq.(\ref{enFourier})
with $dq=0.349$, the value associated to $L=18$, and for  which the stability
properties of the solutions of (\ref{unaecu}) and those of the truncated model
(\ref{enFourier}) are  different. The  same initial value as in the
numerical integration shown in Fig. 6a was given to the modes included in
(\ref{enFourier}). Figure 6b (dotted line) shows that the fastest growing  mode
$q_2$ dominates the final state, in contrast with the numerical integration of
Eq.(\ref{unaecu}).

When Eq.(\ref{enFourier}) is enlarged to include the mode
$q_3=3 dq$, which is linearly stable at $t=0$, the stationary solutions
in which only one modal amplitude is different from zero
are the same as above, but its stability properties
change. For example, the stationary solution dominated by $q_2$ becomes
now unstable for $dq > (10/115)^{1 \over 2} \approx 0.295$  \cite{foot24}.
Numerical solution  of (\ref{enFourier}) including this fourth mode and
keeping $dq=0.349$ shows (dashed line in Fig. 6b) that $\theta_{q_2}$
decays at $t \sim 300 $.
Then, a state dominated by  $q_1$ is approached, as in the integration of
Eq.(\ref{unaecu}) but, in contrast with that continuous model, this is here the
final asymptotic state. In this state $q_3$ is slightly developed and the other
modes have a small amplitude. Again, this is a stable solution of the set of 4
equations and the mode $q=0$ is not yet reached. Figure 6 (solid line)
shows that inclusion
of up to $7$ modes alters the  time scales but not the qualitative picture.

As a conclusion, using only the set of linearly unstable modes, or only some
additional ones, is not enough to describe the time evolution of the continuous
system (which we expect to be well described by the simulations with $72$
grid  points, equivalent to $72$ modes). A large number of linearly
stable  modes are relevant, although its amplitude remains always very
small. It is worth noting that sets of equations containing
the  linearly unstable modes or only a few more are often used with success
in the  literature \cite{Newell,modos3}. The
source of success in these cases is not usually explicitly stated. It becomes
clear from the discussion of the example above that the truncation to the set
of
linearly unstable modes can be useful, at most, when these modes are
associated with stable
stationary solutions of the problem. Otherwise, such truncations can produce
incorrect results.

\acknowledgements

We are indebted to Dr. J. Vi\~nals for many useful suggestions, invaluable
discussions and for his help in solving problems associated with the numerical
procedures. Useful discussions with Dr. P. Coullet and Dr. D. Walgraef are also
acknowledged. This work has been partially supported by Direcci\'on General de
Investigaci\'on Cient\'{\i}fica y T\'ecnica (DGICyT, Spain), under contract
PB 89-0424 and a grant from NATO, within the program "Chaos, Order and
Patterns: Aspects of Nonlinearity", Project No 890482.

\figure{The amplifying factor $\omega(q)$ of equation (\ref{factorw}) in four
different cases. See text for details.}

\figure{Time evolution of $\theta(x) \ $ from a particular initial condition
and $L=256$. The vertical scale is the same for all times. b) Local wavenumber
$q(x)$ for the configuration at $t=40$ of Fig.2a.}

\figure{a) Time evolution of the structure factor averaged over $50$
runs for $L=128$. Times showed are, from bottom to
top: $t=20$, $25$, $30$, $50$, $100$, $200$, $500$ and $750$.
b) Area A(t) of $S(q,t)$ for 4 independent runs ($L=128$).
c) Correlation length $l_c$ as a function of time (see text for details).}

\figure{Number of zeros per unit length in $\theta(x)$
{\sl versus} time for $L=128$ (solid line) and $L=256$ (dashed line). In the
inset the same quantity is plotted versus $\langle q \rangle \ $
(for $L=128$). For $L=128, 256$ an average over 50, 20 initial conditions
respectively was taken.}

\figure{a) Area of the averaged $S(q)$ as a function of time for systems
of different sizes. For size $L=128$ the structure factor has been averaged
over
50 different initial conditions. For the other system sizes the average is over
20 initial conditions. b) The same area scaled by $L^{1/2}$.}

\figure{a) Amplitudes $\theta_{q_n}$ for $n=0,1,2$ and $3$
($\theta_{q_3}$ being the first linearly stable mode) obtained by integrating
equation (\ref{unaecu}) for a particular initial condition of the form
(\ref{inicial}) with $L=18$. b) The same amplitudes obtained from equation
(\ref{enFourier}) (dotted line), from equation (\ref{enFourier}) enlarged
to include mode $q_3$ (dashed line), and including up to mode $q_6$ (solid
line).}


\begin{references}
\bibitem{foot1} Equation (\ref{unaecu}) can be related to the
simplified nematodynamic equations in the form
$\dot \theta = (a_1 - a_2 \partial_x^2)
(b_1 \partial_x^2 \theta + b_2 \theta - b_3 \theta^3)$.
With, $a_1= \gamma_1^{-1}$, $a_2= d^2 (4 \pi^2 \nu_2)^{-1} (1+\lambda)^2$,
$b_1=
K_3$, $b_2= \chi_a H^2 - K_2 (\pi / d)^2$ and $b_3= \chi_a H^2 / 2$. The
meaning
and the typical values of the parameters involved can be found in
\cite{Srajer}.
If a rescaling is performed: $\theta' = b_2 \theta$, $x' = (b_2/b_3)^{1/2} x$
and $t' = b_2^2 a_2 / b_1 t$, equation  (\ref{unaecu}) with c=1 is eventually
obtained with a=0.02 and b=3.0 for an assumed magnetic field of $H = 8 \> kG$.
\bibitem{footsq} Note that the normalization of $S(q,t)$ used in Ref.
\cite{FSS} is different from the one in Eq. (\ref{sq}). For that normalization
$\alpha = d$ in spinodal decomposition problems.
\bibitem{foot24} Note that a value $dq=0.295$ is in the range $dq < 1/3$ in
which there are more than three unstable modes and a description in
terms of only three modes is already {\sl a priori} inadequate.
\end{references}
\end{document}